# REFERENCING TOOL FOR REPUTATION AND TRUST IN WIRELESS SENSOR NETWORKS

Mohammad Abdus Salam[1] and Alfred Sarkodee-Adoo[2]

[1]Department of Computer Science, Southern University, Baton Rouge, Louisiana, USA

[2]Department of Electronics Engineering Technology, Southern University
Baton Rouge, Louisiana, USA

## ABSTRACT

*Presently, there are not many literatures on the characterization of reputation and trust in wireless sensor networks (WSNs) which can be referenced by scientists, researchers and students. Although some research documents include information on reputation and trust, characterization of these features are not adequately covered. In this paper, reputation and trust are divided into various classes or categories and a method of referencing the information is provided. This method used results in providing researchers with a tool that makes it easier to reference these features on reputation and trust in a much easier way than if referencing has to be directed to several uncoordinated resources. Although the outcome of this work proves beneficial to research in the characterization of reputation and trust in WSNs, more work needs to be done in extending the benefits to other network systems.*

## 1. INTRODUCTION

A wireless sensor network (WSN) consists of low cost, low energy sensors, identified as nodes in a WSN, which are scattered in an area on land or under water for the purpose of gathering information on specific events for transmission to a base station for analysis. The WSN has many features such as architecture, algorithms, protocols, energy requirements, transceivers, data storage, security, reputation and trust. Many research works are being carried out on these features. However, there are not much literatures on the characterization of reputation and trust which students, scientists and researchers can readily refer to for detail information.

In this paper, a methodology has been proposed to design a way of providing the needed reference tool which will provide a faster search. The various categories of the reputation and trust have been identified and catalogued into selected reference folders. A table has been created which contains each category together with the corresponding number, abbreviation and description. Referencing the table enables the appropriate folder to be identified. The corresponding folder is then selected for information on the category of reputation and/or trust required.

## 2. REPUTATION AND TRUST

The characterization of reputation and trust-based systems depends on the two features since both are dependent on each other. If a sensor node is trustworthy, then that node is expected to have a good reputation. The method of initialization, observation and distribution in the network produces different results. Initially, every node is considered trustworthy and each node trusts all neighboring nodes. The reputation of every node either increases or decreases during the observation period. Should the reputation decrease, then the node can be regarded as untrustworthy and ignored as a malicious node. The initial reputation value of a node is neutral. Based on the method of observation, there are three main categories of reputation and trust[1]:

1. Observation – observing neighboring nodes for reputation values

- Firsthand systems - depend on firsthand information and direct observation of nodes.
- Secondhand systems - depend on both direct observation and information provided by neighboring nodes.

2. Information symmetry

- Symmetric – all the nodes in the network have access to the same information, using the firsthand or secondhand method.
- Asymmetric – all the nodes in the network do not have access to identical information as in the symmetric method.

3. Centralization

- Centralized - reputation and trust values of all nodes are maintained by one node.
- Distributed - each node maintains the reputation of neighboring nodes within the transmission range.
- Local - all nodes maintain reputation and trust values of neighboring nodes, a method used particularly in static WSNs.
- Global - each node has the reputation and trust values of all the other nodes in the network for both static and mobile nodes.

Many systems now use both firsthand and second-hand systems to update reputation in WSNs. However, it is important to avoid overhead and bottlenecks that may lead to the need for a large memory to store reputation and trust values. Large memories are not part of the WSN architecture.

### 2.1. Reputation

In social and behavioural science, reputation is defined as the opinion of one entity about another. In wireless sensor networks, reputation is defined as a sensor node having the opinion that neighbouring nodes will function as expected with reference to past performance.

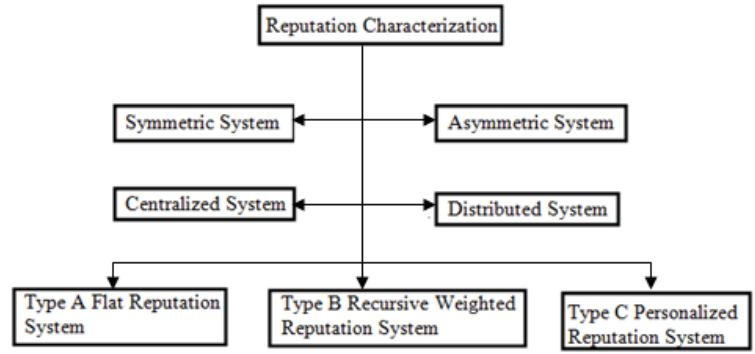

Figure 1. Characterization (Reputation)

With reference to Figure 1, reputation systems within a WSN can be characterized as follows [2]:

- Symmetric systems: the nodes have access to both firsthand and secondhand information.
- Asymmetry systems: the nodes do not have access to firsthand information.
- Centralized systems: reputation is maintained by a single node.
- Distributed systems: each node maintains the reputation of neighboring nodes within the transmission range.
- Type A flat reputation systems: the reputation values are calculated from the trust opinion of the nodes. The opinion of each node has the same weight, whether true or false.
- Type B recursively weighting reputation systems: the reputation values are computed iteratively. Consequently, new reputation values are calculated by using the old weighted values from the previous iteration.
- Type C personalized reputation systems: trust anchors or roots are first selected, which includes the sensor node making the reputation request. This is followed by the opinions of other nodes from previous iterations. The procedure is repeated until all trustworthy nodes in the network are included. The opinions of all untrustworthy nodes are ignored as long as the selected opinions are accurate and trustworthy. Implementation of type C for each node yields different reputation values for the same node [3].

## 2.2. Trust

In social and behavioural science, trust is defined as the feeling that somebody or something can be relied upon or being sure about something even if there is no proof. In wireless sensor networks, trust is defined as the belief that a sensor node is reliable, good and effective based on the reputation of that node.

Trust can be characterized according to how it is implemented. With reference to Figure 2, trust can either be subjective (node trust status at runtime) or objective (actual node trust status) depending on the task being performed.

Trust can be quality of service (QoS) trust which is associated with energy, unselfishness, competence, cooperativeness and reliability. Trust can be classified as computational and can be defined as the trust relation among devices, computers, and networks. Trust can also be direct or indirect. Direct trust deals with direct observations and refers to first hand information. Indirect trust deals with indirect observation and refers to second hand information[4]. A general characterization of trust is shown in Figure 2.

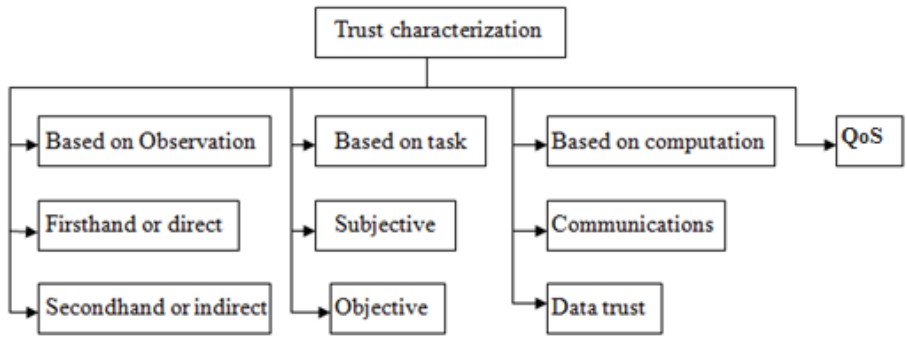

Figure 2. Characterization (Trust) [4]

## 3. CHARACTERIZATION

Designing a scheme for characterization of reputation and trust begins with naming the categories to help identify each class. Several classes have been identified as indicated in Table 1. The first column provides the category number. The second column indicates the abbreviation for the third column, which provides the description of each category. A total of fourteen reputation and trust categories were identified. This number equates to fourteen clearly labelled folders. After identifying the categories, a filing system was designed that will group research papers based on each category which allows for ease of reference. The reference filing system consists of approximately fourteen (14) folders each labelled as indicated in Table 1. After inserting the research papers containing the corresponding categories into the appropriate folders, the following series of steps must be followed to obtain the required information for the specific category:

- Refer to the table containing the numbering scheme, abbreviations and description of the categories.
- Select the category by description.
- Note the abbreviation corresponding to the description.
- Select the folder number matching the abbreviation and description.
- Pull out the numbered folder and start the search for the information required.

Table 1. Categories and folders

| Category Number | Abbreviation | Description |
|---|---|---|
| 1 | OBS | Observation |
| 2 | IS | Information Symmetry |
| 3 | CENT | Centralization |
| 4 | FD | Firsthand or Direct |
| 5 | SI | Secondhand or Indirect |
| 6 | SUB | Subjective |
| 7 | OBJ | Objective |
| 8 | COMP | Computational |
| 9 | COMM | Communication |
| 10 | DT | Data Trust |
| 11 | QOS | Quality of Service |
| 12 | TA | Type A Reputation System |
| 13 | TB | Type B Reputation System |
| 14 | TC | Type C Reputation System |

For example in Figure 3, if you need information on subjective (SUB) classification of trust and reputation, the steps previously referred to lead you to folder number 6. Select folder number 6 and then start the search for the information required in the enclosed papers.

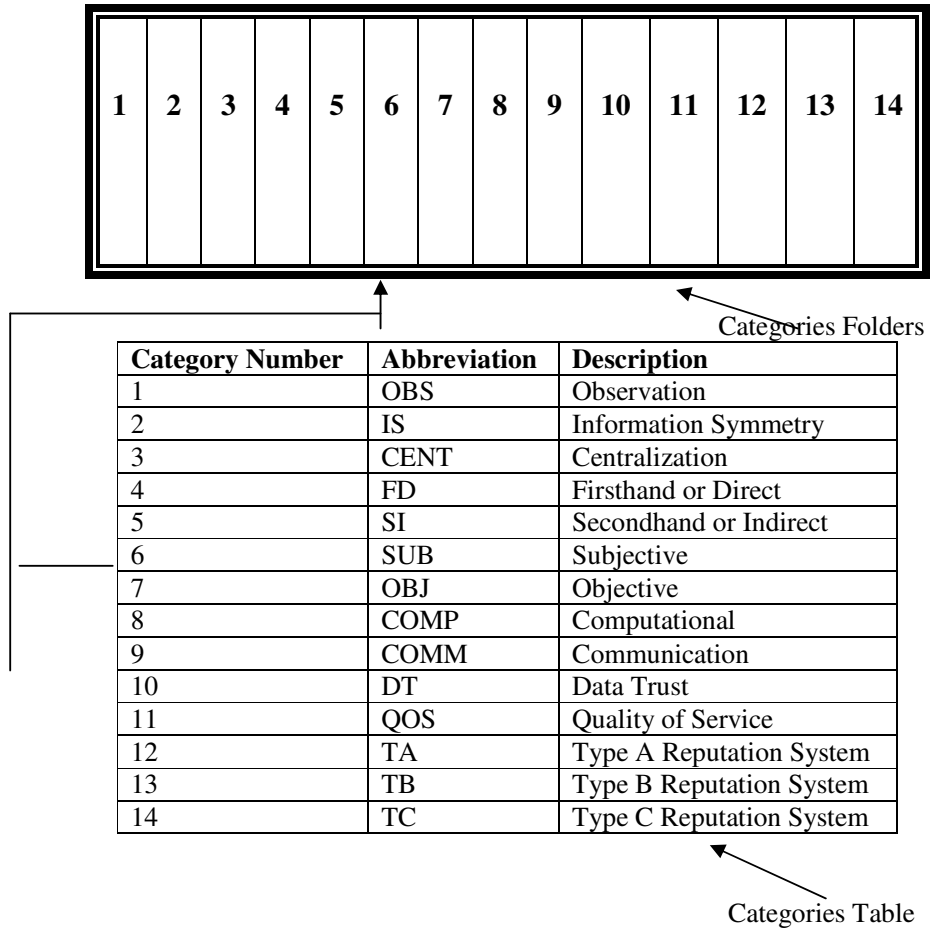

Figure 3. Selection of reference folder

## 4. THE REFERENCE FILING SYSTEM

Sections 4.1. to 4.14. describe the contents of each folder.

### 4.1. OBS Observation folder

Folder number 1, the OBS observation folder, contains papers on observation of sensor nodes in order to determine the reputation of those nodes [4]. V. Geetha and K. Chandrasekaran explain that observation can also help build up trust in a wireless network [6]. Belief level can be user defined trust that is based on direct or indirect observation of node behaviour. Considering that nodes in a wireless sensor network are static, direct observation can be used.

143

### 4.2. IS Information symmetry folder

The IS information symmetry folder contains papers based on information symmetry, which deals with the way information on trust and reputation is accessed. Information symmetry is the condition in which all relevant information is known to all nodes in a particular neighborhood. Information asymmetry on the contrary means some nodes have more relevant information than other nodes. Srinivasan*y* et al. discuss the two modes of symmetry: symmetric and asymmetric mode [1]. In the case of symmetric mode, the nodes can access the same information while with asymmetric mode, the nodes have access to different information.

### 4.3. CENT Centralization folder

The centralization folder (CENT) consists of research papers that contain information on centralization trust [1]. The centralization category consists of two models, mainly centralized and distributed models. In the centralized mode, the reputation and trust values of all nodes are maintained by one node while with the distributed mode, there are two additional modes: local and global. In the case of the local mode, all the nodes maintain the reputation and trust values of neighboring nodes for static WSNs. In the global mode, the nodes have the reputation and trust values of all other nodes for both static and mobile WSNs.

### 4.4. FD Firsthand or direct folder

The FD folder contains papers on firsthand or direct trust. The firsthand system relies on firsthand information and direct observation of the nodes [1]. Firsthand or direct information is more valuable than secondhand information or indirect information. Relying on secondhand information can introduce malicious nodes into the network [4].

### 4.5. SI Secondhand or indirect folder

This folder would contain papers on secondhand or indirect trust. Secondhand refers to information obtained directly from neighboring nodes [4]. Secondhand information is not as reliable as firsthand information. Wu et al. mention that to calculate the secondhand information, the firsthand information is needed as well [7]. Firsthand information is more reliable than that of the secondhand.

### 4.6. SUB Subjective folder

Papers on subjective trust can be found in the SUB folder. Subjective trust is generated as a result of network protocol execution at runtime [4]. When no data transfer is taking place in the network and all nodes are dormant (off or on standby), the trust of the nodes are objective. As soon as a request for data is made by the execution of network protocols and the nodes are turned on, the trust status of the nodes become subjective. R. Christhu et al. state that because of the uncertainty and inaccurate input data that may be encountered in a network, fuzzy logic is used for the evaluation of subjective trust [8].

### 4.7. OBJ Objective folder

All papers on objective trust would be found in the OBJ folder. Objective trust of a node is the trust of a node when the node is dormant [4]. When the nodes in a network are asleep, idle or on standby, then their trust is objective. As soon as a request is made to transmit data (network protocol initiated), then the status changes from objective trust to subjective trust. According to V. Reshmi and M. Sajitar, objective trust obtained globally from the probability model stochastic petri nets (SPN) can be authenticated against the subjective trust derived from execution of the trust management protocol [9].

### 4.8. COMP Computational folder

Computational trust issues could be referenced in the COMP folder. Computational trust is user trust through cryptography (encryption of data), which is based on the authentication of nodes through the use of authentication mechanisms including public key infrastructure (PKI) [4]. According to J. Sabater and C. Sierra, computational trust is used to ensure that an intelligent agent (node) trusts another agent and is able to transfer some responsibilities or tasks (such as information transfer) in a multi-agent environment [10].

### 4.9. COMM Communication folder

This folder would contain papers on communication trust. Geetha et al. believe that, among other trust categories, trust also can be observed as communication trust [6]. The researches define trust as Communication trust can be defined as the belief that nodes in a WSN can maintain connectivity to allow for the transmission and reception of data from node to node virtually error-free [6].

### 4.10. DT Data trust folder

All papers on data trust can be found in the DT folder. Data trust is calculated from sensed data from sensor nodes [11]. Data trust must be considered together with communication trust in order to obtain an accurate trust value. It is therefore inadequate to regard a node as trustworthy only by considering data trust alone. For this reason, the Bayesian fusion algorithm is used in combining the two trusts together to obtain an accurate trust value. Figure 4 illustrates the computation of the true trust value from both the data trust and communication trust [11].

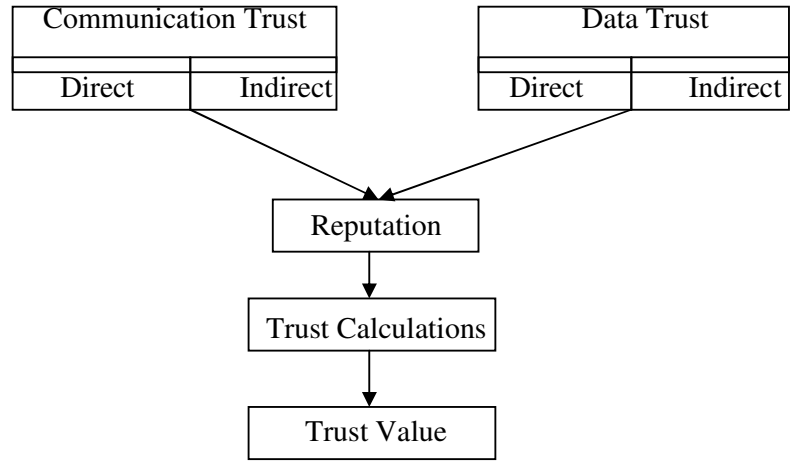

Figure 4. Extended trust computational model for WSNs [11]

### 4.11. QoS Quality of service folder

Papers on quality of service (QoS) trust can be located in the QoS folder. QoS trust is associated with energy, unselfishness, competence, cooperativeness and reliability of sensor nodes [12],

International Journal of Computer Networks & Communications (IJCNC) Vol.7, No.4,July 2015

[13]. Each node is normally powered by batteries. The energy level of the battery determines how long the node will be operational before it reaches a thresh hold value. Beyond the thresh hold value, the node will not be allowed to process data and will be blacklisted and isolated. The unselfishness of a node implies being able to execute all tasks assigned to it without exceptions. This means that a node must accept all information and forward the same to the intended destination.

### 4.12. Type A Reputation system folder

Papers on type A flat reputation system are contained in the Type A folder. With type A flat reputation systems, the reputation values are calculated from the trust opinion of the nodes [3]. The opinion of each node is regarded as equally weighted. Consequently, the final trust value is influenced by the opinion of all the nodes, whether good or bad. Normally, the nodes have to decide whether to express their opinion or not.

### 4.13. Type B Reputation system folder

The type B folder contains information on type B recursively weighting reputation systems. The reputation values are calculated iteratively [3]. The calculated reputation value is improved by increasing the opinion weights of the nodes with higher weights iteratively. The new reputation values of the nodes are calculated from the weighted opinions of the other nodes that were used in the previous iteration.

### 4.14. Type C Reputation system folder

Information in the Type C folder concerns the type C personalized reputation system. Trust roots or anchors are first selected by the node. This selection of the anchors includes the trust requester itself. Initially, the first selections are considered. The opinion of the other roots from trustworthy nodes are then considered, following persistent iterations until final trustworthy nodes are realized. The iteration continues until all trustworthy nodes are selected. The opinion of untrustworthy nodes is ignored as long as the opinions of the trustworthy nodes are correct.

## 5. DATA ANALYSIS

The use of separate folders, each dealing with a single category, makes it easier to provide thorough and in-depth information. The folders also provide a way of upgrading and extending the contents as more research work is carried out on various categories. The information for the categories was obtained from [12]. The following are the descriptions of each category:

### 5.1. Observation

For the observation category, reputation values are obtained by observing neighboring nodes [1]. The observation could be firsthand or secondhand. Node A observes node B over a period of time and builds up a reputation of B. Similarly, node B will observe node A and build up a reputation of A. Alternatively, node A obtains the reputation values from neighboring nodes. Reputation allows nodes to develop confidence in each other so that their behavior in the processing of data is as expected. Any deviation from expected behavior will result in the labeling of that node as a malicious, compromised or faulty and information will no longer be accepted from that node or sent to that node. The offending node will perform no tasks in the network.

## 5.2. Information symmetry

With information symmetry, there are two types of methods to obtain reputation and trust values. The first is the symmetric method, where all nodes obtain the same information from a central point, for example from a cluster head. The second is the asymmetrical method, where all nodes obtain different information from other nodes [1].

## 5.3. Centralization

The centralization category has two options, namely centralized or distributed [1]. With the centralized option, reputation and trust values of all nodes are maintained by one node, while in the case of distributed option, there is the local option where all nodes maintain reputation and trust values of neighbouring nodes, a method used particularly in static WSNs. Also there is the global option where each node has the reputation and trust values of all the other nodes in the network for both static and mobile nodes.

## 5.4 Firsthand or direct trust

Firsthand or direct trust refers to firsthand information or direct observation of a node. The node observes the neighbouring nodes over a period of time and builds up a firsthand trust table from the results [1]. Firsthand trust is more reliable than secondhand trust which is discussed in section 5.5, since the calculated value is based on information obtained directly from neighbouring nodes.

## 5.5 Secondhand or indirect trust

Secondhand or indirect trust refers to information obtained from neighbouring nodes. In this case, the node forwards a request to neighbouring nodes for the trust values needed. Again these values are used to compute the secondhand trust and to build up the trust table created and maintained at each node [1]. There is no guarantee that the trust values received from neighbouring nodes are accurate.

## 5.6. Subjective trust

Subjective trust is the trust of a node during runtime of the network protocol [4]. While the protocol is being executed and there is transfer of information from one node to the other, the trust of the active node is subjective. As soon as the data transfer is completed and the node becomes inactive, the trust reverts to objective trust. Subjective trust can also be the trust status of a node in the "on" state.

## 5.7. Objective trust

Objective trust of a node is the actual trust when the node is inactive [4]. As already explained in section 5.6, this is the status of an inactive node which is not transferring any information in the network at that moment. A node which is in the "off", standby or idle state can be regarded as having objective trust status.

## 5.8. Computational trust

Computational trust is user trust through cryptography (encryption of data), which is based on the authentication of nodes through the use of authentication mechanisms including public key infrastructure [4]. The information being transferred from node to node is encrypted to stop malicious hackers from either corrupting or retrieving the information.

## 5.9. Communication trust

With regards to communication trust, the transmission of information in a network occurs whether communication has occurred or not [8]. The nodes in the network send whatever information is received without ensuring that communication links to the intended destination can relay the information. Good communication links will transmit the entire message and bad links will drop some data packets, resulting in corrupted messages at the base station. A retransmission of the message will then be required.

## 5.10. Data trust

Data trust is a value which depends on the maliciousness, cooperation and competence of a node [15]. When a message is received by a node in the network, it is expected that the node will send the same unaltered message to other nodes in the network. The metric for this trust is referred to as data trust. It is a measure of how trustworthy is the node in handling data without any distortion.

## 5.11. Quality of service trust

Quality of service (QoS) trust is associated with energy, transmission delay, error rate, bandwidth, throughput, competence, cooperativeness and reliability of sensor nodes [12]. The node must have sufficient power supply (energy) to complete transmissions of all messages. Congestion of data in the network must be avoided to minimize delays and dropping of data packets as a result. Bandwidth depends on the data rate within the network. Reliability depends on meantime to failure or repair, meantime between failures and packet/cell loss rate.

## 5.12. Type A, B and C Reputation systems

In the case of type A reputation system, the reputation values are calculated from the trust opinion of the nodes. The opinion of each node has the same effect, whether true or false [3]. With Type B reputation system, the reputation values are computed iteratively. Consequently, new reputation values are calculated from the previous weighted values obtained through iteration [3], [14]. With regards to Type C reputation system, trust reference values are first selected, which includes the sensor node making the reputation request. This is followed by the opinions of other nodes from previous iterations. The procedure is repeated until all trustworthy nodes in the network are included. The opinions of all untrustworthy nodes are ignored as long as the selected opinions are accurate and trustworthy [3], [5].

## 6. STRUCTURE OF THE CHARACTERIZATION OF REPUTATION AND TRUST

The overall structure of the characterization of reputation and trust is illustrated in Figure 5. From the categories table, the description of the trust or reputation is used to obtain the folders from which the research papers containing the trust or reputation information can be found. The categories table is divided into three columns: description, abbreviation and category number. The categories folders section contains fourteen folders, each folder corresponding to the number indicated in the categories table. After identifying the category to search for, the number leads to the exact folder containing the documents on the category selected.

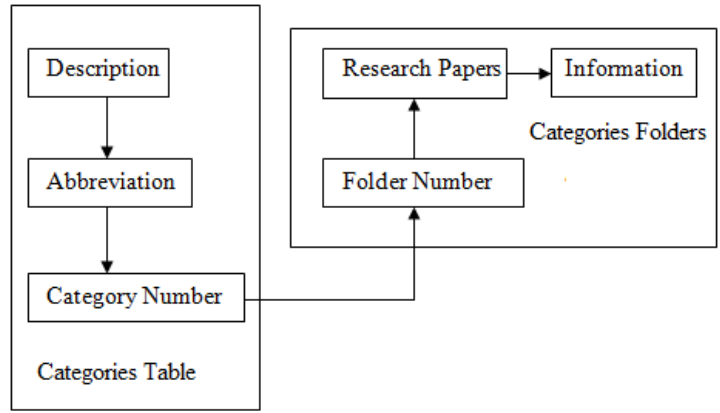

Figure 5. Structure of the Characterization of Reputation and Trust

## 7. CONCLUSION

Reputation and trust are very important features in the wireless sensor network research field. The security of a wireless sensor network is the main concern for those employing WSNs to collect and analyze data at various locations, both on land and under water. Most research papers merely devote a small portion of their findings to this subject. For our research, we have therefore devoted the entire work on the study of various categories of reputation and trust which we believe will enable scientists, researchers and students to have an easier way of accessing information on the various categories of reputation and trust. Having data on reputation and trust will help in designing a network with the best and accurate reputation and trust values so that users will have much needed security in the collection of accurate data in the locations being monitored. The method used in the study will definitely provide a much faster way of obtaining information on the categories of trust to improve deadlines in any design work.

Most of the methods used by authors to explain how to obtain reputation and trust were not identical. However, the methods produced the outcomes expected. It is important to remember that reputation and trust of nodes are obtained firstly by observing the behavior of these nodes for a period of time and then, secondly to generate the metrics or parameters which are then used in computations to obtain the trust values. The results, which are the reputation and trust of the sensor nodes, are then tabulated in tables held by each node. Each category of trust is unique due to their properties. However it is worth noting that some of these categories are linked and in some cases one could be in opposition to the other, depending on which trust value is being applied to the node under consideration.

**Future Work**

More information needs to be extracted from other sources and inserted in the appropriate folders for ease of reference. Compiling information in the folders would be an ongoing process that will continue to supplement the research work being carried out on WSNs globally. We propose that whoever will carry out research on characterization of reputation and trust can obtain an introduction to the subject by referring to the folders for information. This study can be expanded in the following areas:

- Peer-to-Peer (P2P) networks.
- Mobile Ad-Hoc Networks (MANETs).
- Vehicular Ad-Hoc Networks (VANETs).
- Other Wired and Wireless Networks.

## ACKNOWLEDGEMENTS

The authors would like to express their appreciation to those who contributed in any way to the success of the work.

**Authors**

**Mohammad Abdus Salam** is an Associate Professor in the Department of Computer Science at Southern University, Baton Rouge, Louisiana. He received his BS degree in Electrical and Electronics Engineering from Bangladesh Institute of Technology, Rajshahi in 1991 and MS and Ph.D. degrees from Fukui University, Japan, in 1998 and 2001 respectively. Prior to 2005, he worked as an adjunct member of Mathematics and Computer Science at the City University of New York at New York College and as a post-doctorate fellow in the Department of Electrical and Computer Engineering at the University of South Alabama, Mobile, Alabama. He is a senior member of IEEE. His research interests include wireless communication, error control coding and sensor networks

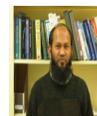

**Alfred Sarkodee-Adoo** earned a Master of Science degree in Computer Science in 2015 at Southern University, Baton Rouge, Louisiana. He completed his BS degree in Electrical and Electronic Engineering at University of Nottingham, Nottingham, United Kingdom in 1971. He worked for the Telecommunications Corporation, as Installations and Maintenance Engineer in Ghana, and also worked for the Volta River Authority as Power Transmission Systems Protection and Control Engineer in Ghana. He is a member of the IEEE. 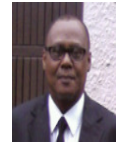